# Generative Machine Learning Models for the Deconvolution of Charge Carrier Dynamics in Organic Photovoltaic Cells

Raymond Li[1*], Flora Salim[1], Sijin Wang[2] and Brendan Wright[2]

[1]School of Computer Science and Engineering, UNSW, High Street, Kensington, Australia
[2]School of Photovoltaic and Renewable Energy, UNSW, High Street, Kensington, Australia
*email: raymond.x.li@unsw.edu.au

**Abstract**

Charge carrier dynamics critically affect the efficiency and stability of organic photovoltaic devices, but they are challenging to model with traditional analytical methods. We introduce **β-Linearly Decoded Latent Ordinary Differential Equations (β-LLODE)**, a machine learning framework that disentangles and reconstructs extraction dynamics from time-resolved charge extraction measurements of P3HT:PCBM cells. This model enables the isolated analysis of the underlying charge carrier behaviour, which was found to be well described by a compressed exponential decay. Furthermore, the learnt interpretable latent space enables simulation, including both interpolation and extrapolation of experimental measurement conditions, offering a predictive tool for solar cell research to support device study and optimisation.

**Introduction**

A detailed understanding of charge carrier dynamics in organic photovoltaic (OPV) devices is critical to optimising for power conversion efficiency and long-term stability, but remains difficult to model due to complex, incompletely understood processes [1]. Traditional analytical frameworks are overly complex or fail to accurately capture the full range of observed behaviours, hindering the development of comprehensive system models and therefore limiting progress in OPV technology optimisation and materials discovery [2].

This study leverages machine learning (ML) to decompose and enable better modelling of charge carrier dynamics. Specifically, we introduce $\beta-$ Linearly-decoded Latent Ordinary Differential Equations ($\beta$-LLODE), an unsupervised model that learns a disentangled latent representation of underlying processes. This model can deconvolve separable components of the experimental data and, from this same representation, generatively simulate measurement trajectories. The model facilitated the attribution of a compressed exponential decay (CED) to describe the dominant component of charge extraction behaviour, and the accurate, predictive simulation of charge behaviour under conditions beyond those experimentally measured.

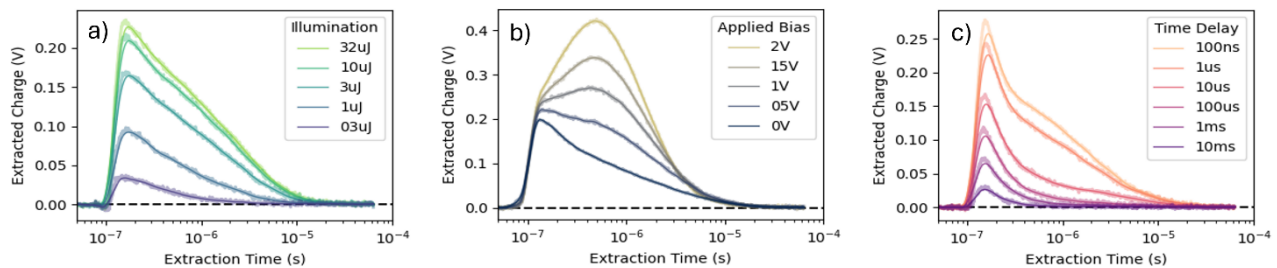

**Figure 1. Examples of measured charge extraction transients (extracted charge over time) for different (a) laser illumination intensities, (b) applied extraction biases, and (c) excitation-extraction time delays, illustrating the complex dynamics involved.**

**Methodology**

Disentangled representation learning was performed by applying the developed $\beta$-LLODE model to a charge extraction dataset, which learnt a decomposition of the measured system dynamics into its constituent components. The transient dataset was experimentally collected for a P3HT:PCBM based bulk heterojunction solar cell using the time-resolved charge extraction (TRCE) technique [2]. This involved exciting the cell with a nanosecond laser pulse, then, after a variable time-delay, extracting the charge carrier population by adding an applied voltage bias. Experimental conditions were systematically varied by adjusting the laser illumination intensity, excitation–extraction delay, and extraction voltage. The dark response, measured without laser illumination but with an applied extraction voltage, was subtracted from the corresponding illuminated measurements. This removes high-impedance switch noise and circuit resistor-capacitor (RC) response, isolating the signal solely due to the photogenerated charge carriers. Each transient represents 50 repeated sampling measurements acquired and averaged to suppress measurement noise. The original 100,000 sample transients were resampled logarithmically for further analysis and model training. The dataset was sparse, with each unique combination of experimental conditions having only one sample, and relatively small in the context of ML datasets, with only 150 data samples in total.

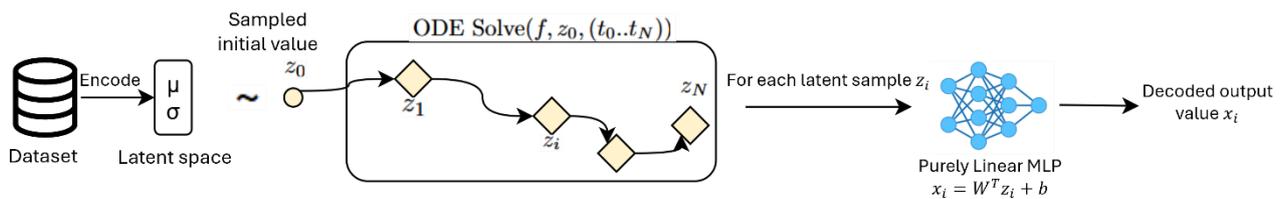

Figure 2. Diagram of the ML model's architecture, including encoder/dynamics/decoder stages. When learning with a $\beta$-VAE objective function, the RNN encoder and ODE solver produce disentangled latent trajectories. At the same time, the interpretable linear decoder is used to decode the latent trajectory back into a charge extraction transient at the model's output.

The ML model ($\beta$-LLODE) combines a latent-ODE [3] backbone with a linear decoder, using a $\beta$Variational Autoencoder (VAE) [4] objective function. As such, the model learnt to reproduce the dataset generatively. The latent-ODE architecture integrates recurrent neural networks (RNNs) with ordinary differential equations (ODEs), where the RNN is used as the encoder in a VAE framework, followed by an ODE solver as the global dynamics function, and finally a linear map as the decoder. This design was chosen because (i) it efficiently encodes time-series data into a compact latent representation, and (ii) the decoding process uniquely samples from the latent space and evolves the resulting state through time, producing smooth, continuous latent trajectories.

A β-VAE objective was used to regularise the latent space, promoting compact, disentangled representations where each latent trajectory corresponded to a distinct component of charge behaviour [5]. Finally, a linear decoder, implemented as a multi-layer perceptron (MLP), maps the latent trajectories from the latent-ODE into the reconstructed transient. A linear decoder introduced interpretability between the latent space and the reconstructed output. A close match between the reconstructed and experimentally measured transients suggests that the learned latent trajectories provide a physically meaningful decomposition of the original data. The produced latent states are then independently multiplied by their respective weights that are obtained through inspection from the linear decoder. This produced trajectories in the output space that decomposed the original transient. During training, we scale $\beta$ to progressively enforce disentanglement as the model learns to reconstruct the data. This scaling $\beta$ value initialised at 0.5 and doubled every 100 epochs, with a maximum value of 4. The 'Adam' optimiser was used with a decreasing learning rate, beginning at $10^{-3}$ and halving every 100 epochs, with a lower limit of $10^{-7}$. Training was observed to converge by 1000 epochs, where a vector of maximum 5 latent dimensions was made available to the model for representation.

**Results**

The model achieved a close fit to the original dataset, with an average mean square error (MSE) of 0.57. Figure 3(a) shows a sample of experimental measurements and the corresponding model predictions, illustrating this very close alignment.

Figure 3(b) shows a decomposition of a sample transient into its disentangled constituent components. The β-VAE objective encouraged compact, efficient representations, identifying two principal features in the charge extraction dataset. Although five latent dimensions were available, the converged model utilised only two as necessary. On an observational level, latent dimension 0 described the mobile photogenerated charge population during extraction, while latent dimension 1 mapped the influence of applied bias on the rate of charge extraction.

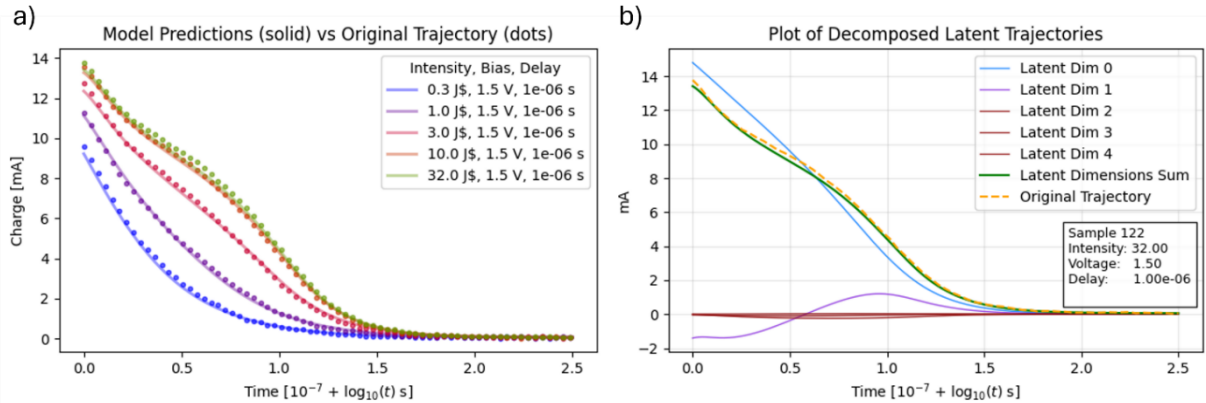

**Figure 3. Example plots showing modelling results: (a) displays a representative sample of raw measurement transients (dots) and corresponding model reconstructions (solid) illustrating the fit quality achieved, and (b) shows the decomposition of a single charge transient into underlying components coupled to latent dimensions.**

The disentangled constituent component represented in latent dimension 0, which isolated the primary charge extraction decay behaviour of interest, was found to better facilitate the fit of mathematical models. In particular, it was found that a CED accurately modelled the primary charge extraction decay behaviour. A CED models a value $y$ as a function of time $t$ by:

$$y = Ae^{-\left(\frac{t}{\tau}\right)^k} + C,$$

where $K \geq 1$ and parameters $A, K, \tau$ and $C$ are tuneable. $C$ was fixed to match the transient convergence value of 0. As shown in Figure 4, fitting the CED to this isolated latent trajectory yielded a markedly more precise fit, reducing the MSE to approximately 11% of that obtained when fitting the original transient.

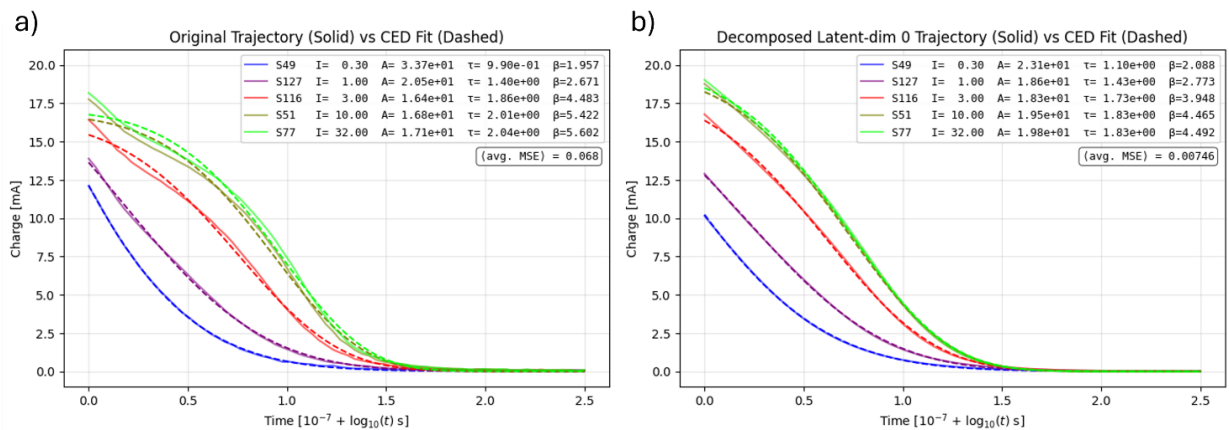

**Figure 4. Comparison of compressed exponential decay (CED) fits: (a) applied to the original charge extraction trajectory, and (b) applied to the isolated decay component extracted from the latent dimensions, illustrating that the latent-space representation enables a substantially improved fit quality.**

CEDs can be expressed as continuous sums of Gaussian distributions, whose parameters (mean, maximum, width, shape) can provide deeper insight [6]. As OPVs are energetically disordered systems, the presence of a CED is expected. The Gaussian distributions arising from the CED are hypothesised to represent the distribution of energy states of various charge populations within the OPV active layer, offering a potential pathway to investigating fundamental charge generation and extraction mechanisms.

The latent vector was found to provide a compact, interpretable representation of OPV charge extraction dynamics, enabling direct simulation within the latent space. By interpolating or extrapolating between latent states derived from experimental data, virtual trajectories could be generated to predict system responses under different experimental conditions (as shown in Figure 5). This approach enables efficient probing of charge-extraction behaviour across a range of operating regimes, facilitating hypothesis testing, optimisation of extraction efficiency, and identification of measurement artefacts with significantly fewer physical trials. This means, in simple terms, that we can accurately predict the results of TRCE measurements given any new set of experimental conditions, enabling a greater investigation of system behaviour without the additional experimental burden.

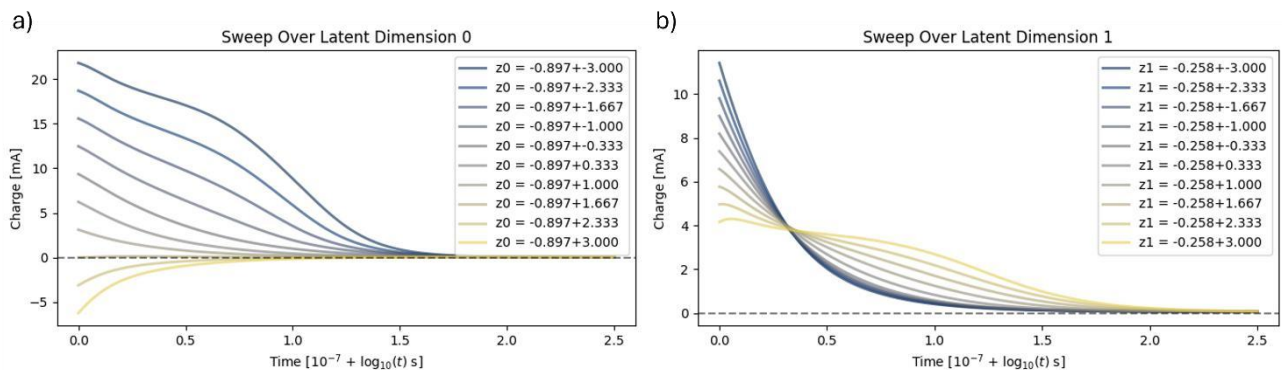

**Figure 5. Sweep over latent dimensions, illustrating correlations of latent dimension 0 (a) with total charge and latent dimension 1 (b) with the rate of charge extraction under an applied bias. The latent space was found to be smoothly structured and interpretable, enabling the controlled simulation of distinct scenarios.**

### Conclusions

This study employed ML models to deconvolute the complex charge extraction behaviours observed in TRCE measurements of OPV devices. The developed approach enables accurate mathematical modelling, revealing that compressed exponential decays provide an excellent fit for the observed charge extraction dynamics after deconvolution and isolation. Additionally, the model enables the simulation of system behaviours through its structured and interpretable latent space. These findings open new avenues for more precise analysis and predictive modelling of OPV systems. Beyond OPVs, the methodology is broadly applicable to other domains involving complex time-series data, offering a versatile tool for uncovering and characterising latent dynamical processes in diverse physical and engineering systems.

**References**

[1] Rahman, M. U. *et al*, **2024**, 'Recent advances in stabilizing the organic solar cells'. *MRS Energy & Sustainability*, vol. 11, pp. 1–20.

[2] Wright, B. F., **2017**, 'Quantifying Recombination Losses during Charge Extraction in Bulk Heterojunction Solar Cells Using a Modified Charge Extraction Technique', *Advanced Energy Materials,* vol. 7.

[3] Higgins, I. *et al*, **2017**, 'beta-VAE: Learning Basic Visual Concepts with a Constrained Variational Framework' *Proceedings of the 5th International Conference on Learning Representations (ICLR).*

[4] Rubanova, Y. *et al*, **2019**, 'Latent ODEs for irregularly-sampled time series', *CoRR*, vol. abs/1907.03907.

[5] Wang, X. *et al*, **2024**, 'Disentangled Representation Learning', *IEEE Transactions on Pattern Analysis and Machine Intelligence*, June.

[6] Hansen, E. W. *et al*, **2013**, 'Compressed Exponential Response Function Arising From a Continuous Distribution of Gaussian Decays – Distribution Characteristics', *Macromolecular Chemistry and Physics*, vol. 214, no. 7, pp. 844–852.